# Optimized Automated Cardiac MR Scar Quantification with GAN-Based Data Augmentation

**Didier R.P.R.M. Lustermans[1], Sina Amirrajab[1], Mitko Veta[1], Marcel Breeuwer[1,2], Cian M. Scannell[3*]**

[1]Department of Biomedical Engineering, Eindhoven University of Technology, Eindhoven, The Netherlands

[2]Department of MR R&D – Clinical Science, Philips Healthcare, Best, The Netherlands

[3]School of Biomedical Engineering and Imaging Sciences, King's College London, London, United Kingdom

**\* Correspondence:**
Cian M. Scannell
cian.scannell@kcl.ac.uk

**Abstract**

**Background:** The clinical utility of late gadolinium enhancement (LGE) cardiac MRI is limited by the lack of standardization, and time-consuming postprocessing. In this work, we tested the hypothesis that a cascaded deep learning pipeline trained with augmentation by synthetically generated data would improve model accuracy and robustness for automated scar quantification.

**Methods:** A cascaded pipeline consisting of three consecutive neural networks is proposed, starting with a bounding box regression network to identify a region of interest around the left ventricular (LV) myocardium. Two further nnU-Net models are then used to segment the myocardium and, if present, scar. The models were trained on the data from the EMIDEC challenge, supplemented with an extensive synthetic dataset generated with a conditional GAN.

**Results:** The cascaded pipeline significantly outperformed a single nnU-Net directly segmenting both the myocardium (mean Dice similarity coefficient (DSC) (standard deviation (SD)): 0.84 (0.09) vs 0.63 (0.20), p < 0.01) and scar (DSC: 0.72 (0.34) vs 0.46 (0.39), p < 0.01) on a per-slice level. The inclusion of the synthetic data as data augmentation during training improved the scar segmentation DSC by 0.06 (p < 0.01). The mean DSC per-subject on the challenge test set, for the cascaded pipeline augmented by synthetic generated data, was 0.86 (0.03) and 0.67 (0.29) for myocardium and scar, respectively.

**Conclusion:** A cascaded deep learning-based pipeline trained with augmentation by synthetically generated data leads to myocardium and scar segmentations that are similar to the manual operator, and outperforms direct segmentation without the synthetic images.

## 1    Introduction

Late gadolinium enhancement (LGE) cardiac MRI is the reference standard for the non-invasive assessment of myocardial viability, and is widely used in clinical routine (1). It has been shown to accurately identify areas of myocardial infarction (2), and the size and transmurality of scar regions are important parameters to guide the management of patients (3). Visual reporting of such parameters is user-dependent, and thus, the robust and accurate quantification of scar would be highly beneficial. If the quantification could be reliably performed automatically, it could also facilitate further adoption of LGE cardiac MRI in clinical practice, particularly in less specialized, low-volume centers.

The standard approach to the quantification of LGE has been the use of a fixed intensity threshold value, usually relative to a reference region. The most common approaches segment scar as being $n$ (typically 5) standard deviations (nSD) above the mean intensity of a remote normal myocardium region or above half of the maximum value of a scar region (full width at half maximum (FWHM)). To date, quantification has primarily been performed in research studies due to the time-consuming manual interaction and lack of reproducibility between operators (4). Advanced methods for the thresholding of scar regions, that do not require manually drawn reference regions, such as Otsu thresholding (5), or fitting to expected distributions using expectation-maximization (6,7) have also been proposed without achieving clinical adoption. In general, these intensity-based thresholding methods are subject to false positives due to noise and imaging artifacts, and they do not incorporate any spatial context in the thresholding.

More recently, deep learning, using convolutional neural networks (CNNs), has become the state-of-the-art for cardiac MRI segmentation in a wide range of applications (8–17), and it has also been applied to LGE segmentation. Fahmy et al. demonstrated accurate scar volume quantification using a 3D U-Net in patients with hypertrophic cardiomyopathy (18) and Zabihollahy et al. proposed a cascaded pipeline which segments the left ventricle (LV) myocardium and scar in two steps using images from multiple planes (19). There has been further interest in the topic as a result of the open-source dataset made available as part of the automatic Evaluation of Myocardial Infarction from Delayed-Enhancement Cardiac MRI (EMIDEC) challenge (20), with several authors investigating the use of cascaded pipelines (21,22) or the incorporation of prior information (23) to improve reliability.

In this work, we developed and evaluated a cascaded deep learning pipeline for automatic myocardium and scar segmentation and quantification in subjects with suspected acute myocardial infarction. In particular, we proposed a cascaded deep learning pipeline, consisting of bounding box detection and myocardium segmentation, followed by scar segmentation, and we investigated the impact of each individual step on the performance pipeline. We further assessed the benefit of including synthetic images, generated by a conditional generative adversarial network (GAN), in the training data (in addition to conventional data augmentation). Both the GAN-based synthetic data and the step of splitting the task into simpler sub-problems are designed to overcome the challenge of the limited amount of training data, and it is hypothesized that both will lead to improved performance with the small amount of training data.

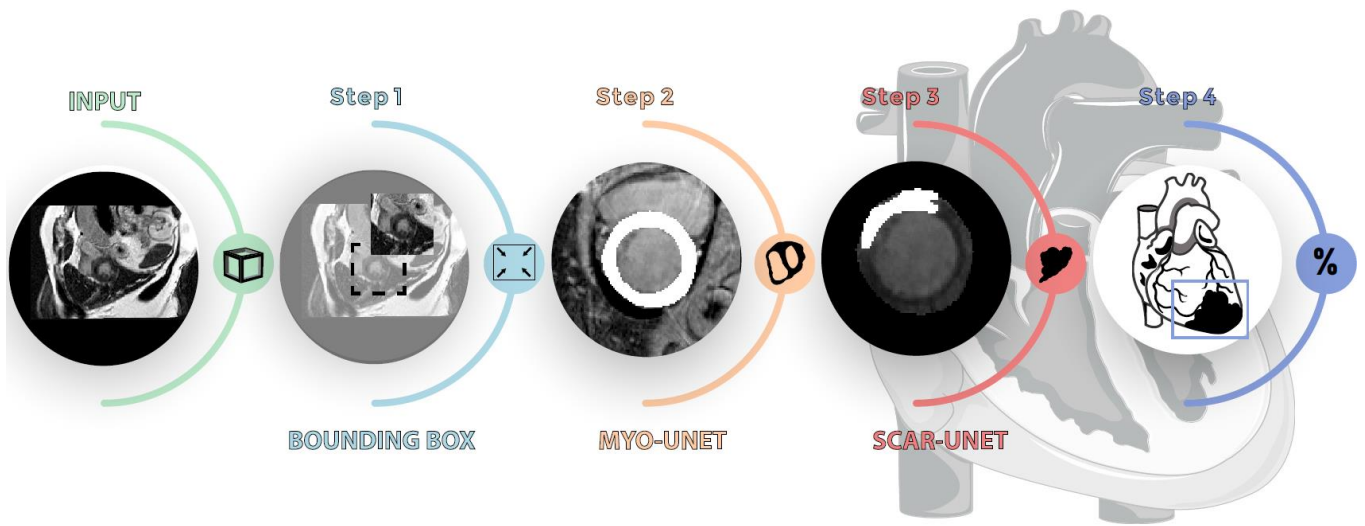

**Figure 1:** *The proposed scar quantification pipeline. Firstly, a bounding box is detected around the heart, followed by myocardium segmentation. Subsequently, scar is segmented, if present, and used to compute the scar burden as a percentage of myocardial volume.*

## 2    Materials and Methods

### 2.1    Dataset

The dataset from the EMIDEC challenge was used (20). This consists of the LGE cardiac MRI scans of 150 patients, out of which 105 are pathological and 45 are normal. The data are divided into training (n=100) and testing (n=50) sets by the challenge organizers. The data acquisition was performed at the University Hospital of Dijon (France) on 1.5T and 3T systems (Siemens Medical Solution, Erlangen, Germany) with a T1-weighted phase sensitive inversion recovery (PSIR) sequence (TR = 3.5 ms, TE = 1.42 ms, TI = 400 ms, flip angle = 20), performed 10 minutes after the administration of a gadolinium-based contrast agent (Gd-DTPA; Magnevist, Schering- AG, Berlin, Germany), at concentration between 0.1 and 0.2 mmol/kg, during a breath-hold. Further details can be found in (20). The images were manually segmented, in consensus, by two expert operators (a cardiologist with 10 years' experience in cardiac MRI and a physicist with 20 years' experience)**.** For the purpose of this work, the scar and microvascular obstruction (MVO) segmentations are combined in one class label representing the total infarction area.

### 2.2    Cascaded Pipeline

As shown in Figure 1, the proposed cascaded pipeline consists of three main steps: 1) the detection of a bounding box that encompasses the LV cavity and LV myocardium, 2) the segmentation of the myocardium, and 3) the segmentation of scar. In this work, deep learning models are trained sequentially to achieve each of these steps. An ablation study is performed by removing steps in the pipeline to analyze their impact of the performance, and the model performance with the inclusion of the synthetically generated images is compared with that of a model without this data augmentation. The segmentation and computed volumes from the automated analysis are compared to the manual quantification on the EMIDEC challenge test set. The trained models for both segmentation and synthetic data generator are made available at https://github.com/cianmscannell/lge-quant-emidec, along with the generated synthetic data.

### 2.2.1 Bounding Box

The bounding box algorithm used is as proposed in Scannell et al. (12). This first assumes that there is a fixed bounding box in the center of the image. A CNN is trained to predict, from a LGE image, the transformation of this proposed bounding box so that it covers the LV myocardium and cavity of the image. This is framed as a regression problem to predict four continuous values, the 2D translation of the center of the box and the scaling of the two different sides of the rectangular box. The proposed bounding box is of size $134 \times 134$ pixels, which is the mean size present in the training set. Due to the shape of the LV, it is sufficient to predict the bounding box on a slice in a basal location. This work uses the second 2D image from the top of the stack to avoid images where the LV cavity and myocardium are not present.

The CNN takes the original images, center-cropped or zero-padded to a size of $256 \times 256$ pixels, as input. These input images are min-max normalized, using the $5^{th}$ and $95^{th}$ percentile of intensity values as the pseudo-minimum and maximum, respectively. The architecture consists of four convolutional layers, each layer with two convolutions using $3 \times 3$ kernels followed by $2 \times 2$ max-pooling. These layers are followed by fully connected layers. Each layer uses batch normalization and rectified linear unit (ReLU) activations, except the output layer which uses a linear activation. A batch size of 32 and L2 regularization on the convolutional kernel parameters was used with a weight of 0.0001. The loss function, the mean squared error, was optimized with the use of an Adam optimizer and convergence was determined by early stopping. Data augmentation was used, consisting of random combinations of rotation, translation, blurring, scaling and noise added to the images. The parameters of the data augmentation are provided in Supplementary Table 1.

### 2.2.2 Myocardium Segmentation

The segmentation of the myocardium is based on a nnU-Net model, as described by Isensee et al. (24). Briefly, the algorithm automatically optimizes the architecture and hyperparameters of a U-Net model based on a fingerprint of the training dataset. A 2D network is used with leaky ReLU activations and instance normalization. The initial number of feature maps was 32 and doubled each layer until it reached a size of 512. Stochastic gradient descent with Nesterov momentum ($\mu$=0.99) is used as the optimizer with an initial learning rate of 0.01. The loss function is the sum between the cross-entropy and dice loss . After a fixed amount of 1000 epochs, the network with the best validation set performance is chosen. The algorithm also includes on-the-fly data augmentation. The input is the cropped LGE image which is reshaped to $128 \times 128$ pixels, and normalized in the same way as described for the bounding box.

Quality control, of the myocardium segmentations is performed in which connected component analysis was used to identify failed (not closed) myocardium segmentations, which were then re-segmented with ten different augmented (by translation) bounding boxes. The 10 predictions are summed and the pixels that are predicted in the myocardium greater than $k$ times are included in the final prediction, where $k$ is the smallest number that yields a closed myocardium segmentation. If no closed myocardium is achieved in this manner, the original prediction is used.

### 2.2.3 Scar Segmentation

A second nnU-Net model is used for the scar segmentation. The network input is a $64 \times 64$ pixel image, cropped according to the contour gravity center of the myocardium segmentation. The image is also masked with the myocardium segmentation to set intensity values outside the myocardium or LV cavity to 0. The myocardium values are normalized to a signal intensity range between 0 and 1, using the 5th and 95th percentiles, as before. The LV cavity is set to a fixed signal intensity value of 2.5. A further quality control step is used that removes small regions of predicted scar by removing regions that lead to a predicted scar-to-myocardium volume ratio less than 3%, as scar sizes smaller than this are not feasible in this population.

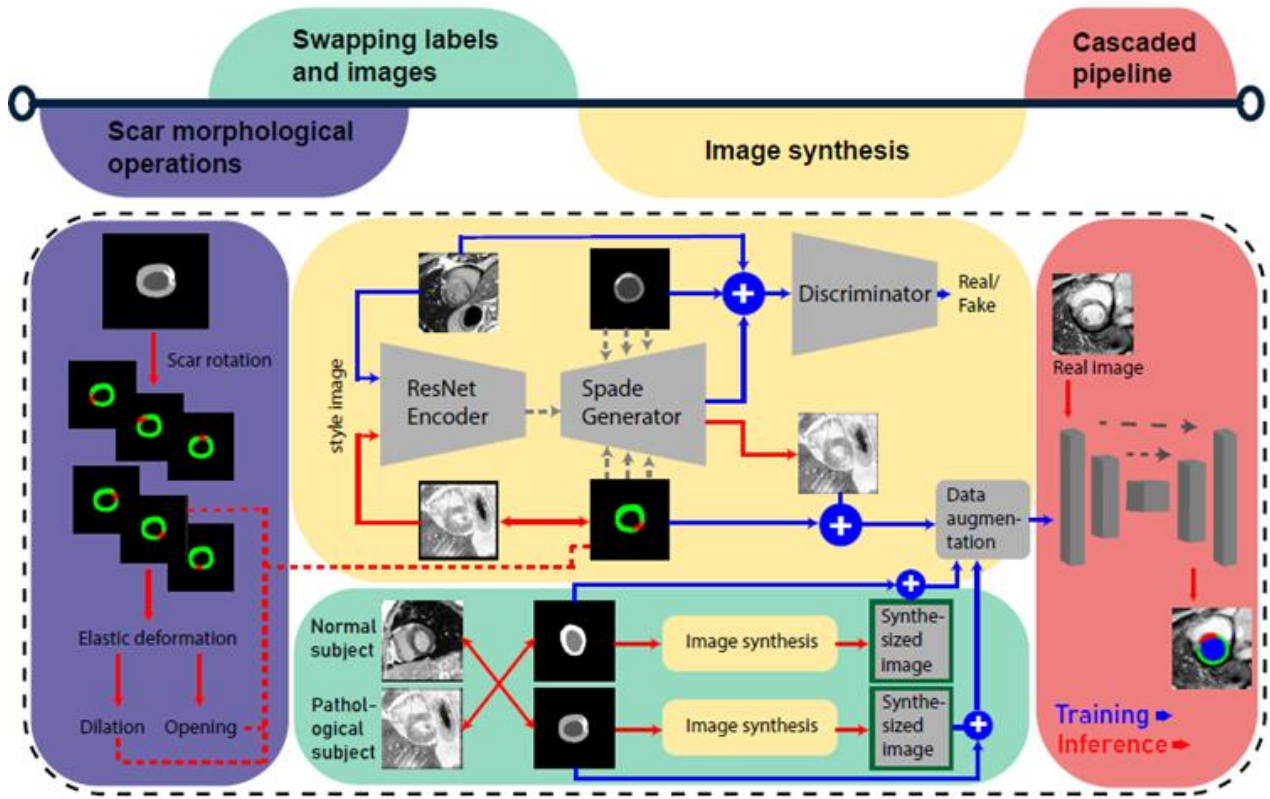

***Figure 2:*** *A flowchart of the synthetic image generation process. For inference, augmented segmentation labels are input to a trained conditional GAN, accompanied by a style image (red). The label maps are generated by swapping existing labels between pathological and normal subjects (green) and by performing morphological operations on scar labels (elastic deformation, rotation and dilation or opening) (purple). The generated synthetic data are then used as augmentation data for the cascaded pipeline.*

### 2.3  GAN-Based Image Synthesis

The image synthesis module, as shown in Figure 2, is based on a ResNet-encoder coupled with a segmentation-conditioned GAN that uses SPatially-Adaptive (DE)normalization layers (SPADE) (25) throughout the generator architecture. The use of SPADE-based conditional GANs for preserving the anatomy of the segmentations for cine cardiac MR image synthesis was investigated previously by Abbasi-Sureshjani et al. (26) and Amirrajab et al. (27) and it was shown that providing multi-class labels to guide the SPADE generator allows the synthesis of realistic images. In contrast to previous works, a LGE image was input to the ResNet-encoder to extract the style and background anatomical information in the synthesis process. This was necessary as the segmentations for this data were only available for the left ventricle and myocardium, so there was no information from the background for the generator to use to synthesize realistic anatomies around the heart. The training process used pairs of LGE style images with the corresponding ground-truth segmentations, and is described in more details in (27).

After training, to generate new pairs of synthetic LGE images with the trained generator, two strategies are used: 1) augmented labels, and 2) swapped labels. For the augmented labels, the segmentation labels from the training set are augmented, by rotation and morphological operations (parameters defined in Supplementary Table 2), to create previously unseen shapes and positioning of scar, and input to the generator (shown in the purple box of Figure 2), and for the swapped labels, existing segmentation labels from pathological patients are combined with style images from normal patients, and vice versa, to create new patients (shown in the green box of Figure 2). The augmentation of the scar segmentations included rotation by a multiple of 60°, elastic deformation, dilation and opening. The augmented and swapped labels are used to generate synthetic data for the myocardium segmentation training, and only the augmented labels are used for the scar segmentation as, due to the masking of the myocardium, changing the background (via the swapped style image) will have no effect.

## 2.4   Evaluation

An ablation study was first performed, trained only with the real patient data, to analyze the impact of the individual steps of the cascaded pipeline, by comparing the myocardium and scar segmentation of (a) the full cascaded pipeline to (b) myocardium and scar segmentation in two steps, without the bounding box, (c) directly segmenting the myocardium and scar with one nnU-Net, with the bounding box, and (d) directly segmenting the myocardium and scar with one nnU-Net, without the bounding box. Supplementary Figure 1 shows a representation of these four methods. As the cascaded nature of the pipeline can lead to the propagation of errors, when an incorrect myocardium segmentation is used to mask the input to the scar segmentation model, this effect was also studied. In particular, the increase in performance found when replacing the predicted myocardium segmentation with the ground-truth in the cascaded pipeline was analyzed. Secondly, the performance was tested by comparing the cascaded pipeline trained only with the real data (a) and a version with training augmented by synthetic images (e), to assess the impact of adding synthetic data to the real dataset. In order to avoid the possibly confounding effect of the cascaded pipeline, a direct one-step segmentation model was also trained with and without the synthetic data augmentation.

These comparisons were performed on a randomly selected internal testing set (N=20) split from the training set. This analysis was performed using the Dice similarity coefficient (DSC) metric on a per-slice level and the performance of models are compared with the Wilcoxon signed-rank test. The final evaluation was performed with the full cascaded pipeline, augmented with the GAN-based synthetic data, the model with the best performance on the internal test set, on the 50 test subjects from the EMIDEC challenge, where the mean DSC, Hausdorff distance (HD) and volume difference per-subject was reported. The automated scar quantification was further evaluated with respect to the manual quantification using Pearson correlation and Bland-Altman analyses. The segmentation labels for the test set are not available publicly but the evaluation of these metrics was performed by the challenge organizers.

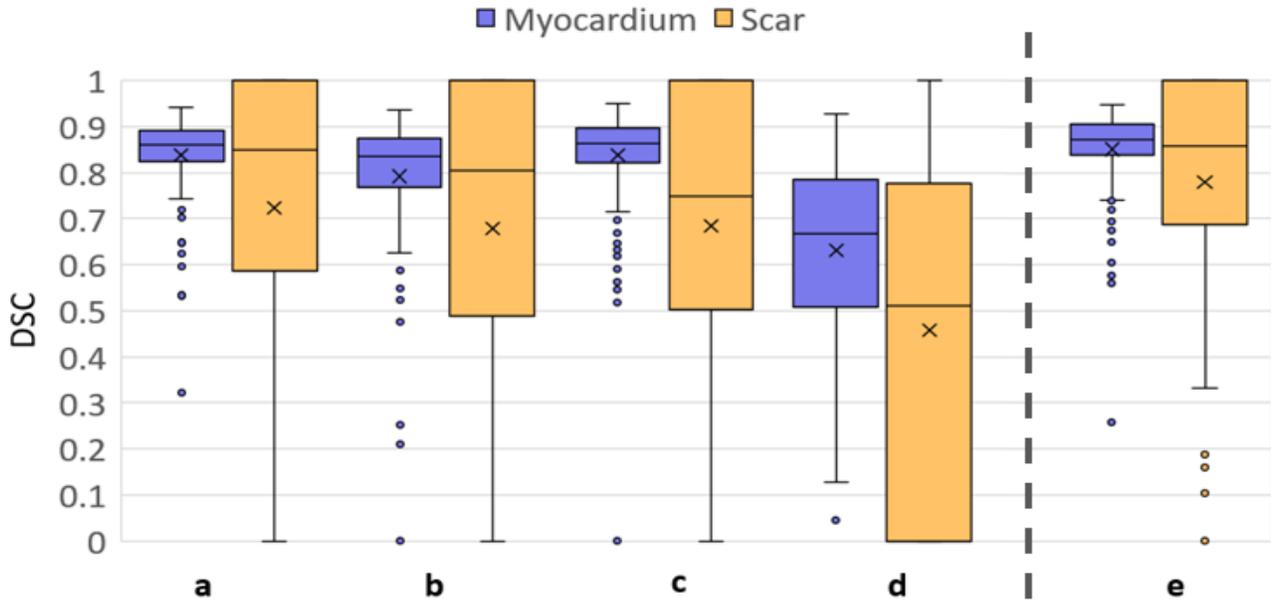

***Figure 3:*** *The distributions of the DSC values for myocardium (purple) and scar (yellow) segmentations on the internal test set, for the five trained versions of the cascaded pipeline. These are (option (a)) the proposed full cascaded pipeline, (b) myocardium and scar segmentation in two steps, without the bounding box, (c) directly segmenting the myocardium and scar with one nnU-Net, with the bounding box, (d) directly segmenting the myocardium and scar with one nnU-Net, without the bounding box, and (e) the full cascaded pipeline trained with the synthetic data augmentation. The X mark indicates the mean value.*

## 3    Results

### 3.1    Cascaded Pipeline

The DSC values for both myocardium and scar between the manual and automatic segmentations, for the four versions of the pipeline on the internal test set, are shown in Figure 3 (a)-(d). The proposed full cascaded pipeline (option (a)) had the highest DSC for both the myocardium (mean DSC (standard deviation (SD)): 0.84 (0.09)  and scar (0.72 (0.34)) segmentations. This was significantly higher than all other approaches including a single nnU-Net directly segmenting both the myocardium and scar (option (d): myocardium DSC: 0.63 (0.20), p < 0.01 and scar DSC: 0.46 (0.39), p < 0.01), the direct myocardium segmentation followed by scar segmentation (option (b): myocardium DSC: 0.79 (0.15), p < 0.01 and scar DSC: 0.68 (0.37), p = 0.02), and the two-step pipelines of bounding box followed by direct myocardium and scar segmentation (option (c): myocardium DSC: 0.80 (0.12), p < 0.01 and scar DSC: 0.68 (0.35), p < 0.01).

### 3.2    GAN-Based Synthetic Image Data Augmentation

Figure 4 shows GAN-based synthetic images generated from one real patient by rotating, elastically deforming, and morphologically opening the original segmentation mask. The DSC values for the full cascaded pipeline trained with the synthetic data are shown in Figure 3 (e). The mean (SD) DSC between the manual and automatic myocardium segmentation increased by 0.01 (from 0.84 (0.09) to 0.85 (0.09)) with the addition of the synthetic training data on the internal test set. The inclusion of the synthetic data for training resulted in a 0.06 increase in the scar DSC (from 0.72 (0.34) to 0.78 (0.28)), a statistically significant difference (p < 0.01), as well as a decrease in the SD. Moreover, both models identified pathological and normal subjects with an accuracy of 100% (20 out of 20) on the internal test set. For the direct one-step segmentation the DSC is also increased from 0.80 (0.12) to 0.84 (0.11) for the myocardium and 0.68 (0.35) to 0.71 (0.33) for scar (both p < 0.01). Replacing all the predicted myocardium segmentations with the manual ground-truth label only gives a modest improvement in DSC for the scar segmentation from 0.78 to 0.80, also on the internal test set.

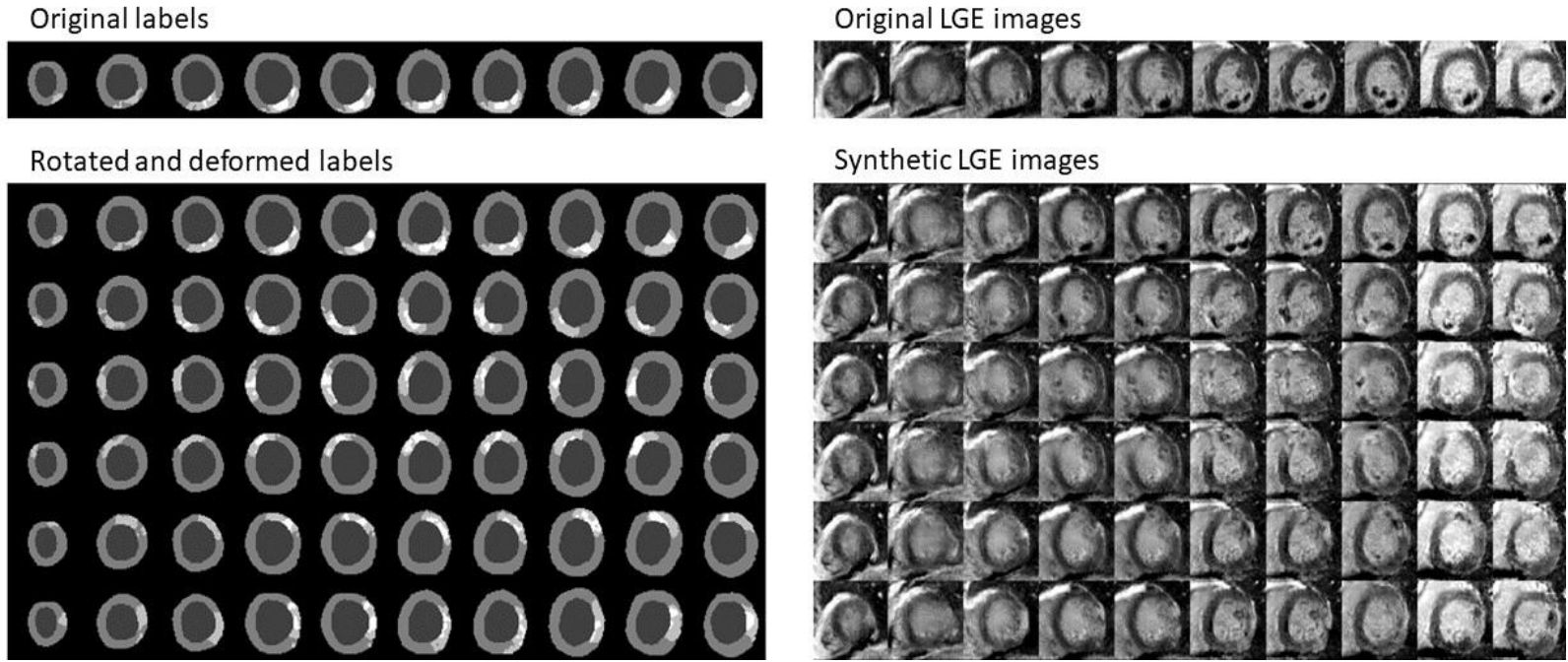

**Figure 4:** *The GAN-based synthetic LGE images generated with a set of rotated and deformed input labels from a single patient (original LGE images and labels shown in the top row) for all slices apex (left) to base (right).*

### 3.3   Challenge Test Set

The proposed cascaded pipeline trained with synthetic data augmentation was evaluated on the EMIDEC challenge test set (N=50 subjects). Figure 5 shows segmentations for two representative patients from the test set showing both good and bad performance (note that the ground-truth segmentations for the test set are not available for comparison). The mean (SD) DSC, HD and volume difference per-subject was 0.86 (0.03), 15.7 (11.9) mm and 11.5 (8.4) cm$^3$ respectively for myocardium segmentation. 5 out of 358 myocardium segmentations failed to generate a closed shape and were identified by the quality control procedure, and a correction was attempted. Secondly, the model showed a mean (SD) DSC and volume difference between the manual and automatic scar regions of 0.67 (0.29) and 41.0 (5.8) cm$^3$ per-subject, and the difference in scar volume relative to the volume of the myocardium was 3.41% (4.8%). Furthermore, the model classified patients as scarred or not with an accuracy of 94% (47 out of 50 subjects).

Figure 6 compares the computed volumes for the automated and manual segmentations in a scatterplot, the Pearson correlation coefficient is 0.96 and 0.94 for myocardium and scar, (a) and (b) respectively. Per-subject, Bland-Altman analysis on the LV myocardial volume (c) showed an agreement between the manual and automatic quantified volumes of the proposed cascaded pipeline with bias of 10.24 cm$^3$ and limit of agreement 19.67 cm$^3$. Additionally, a good agreement between the manual and the proposed automatic quantified scar volume with a bias of 2.74 cm$^3$ and limit of agreement of 13.06 cm$^3$ was shown (d).

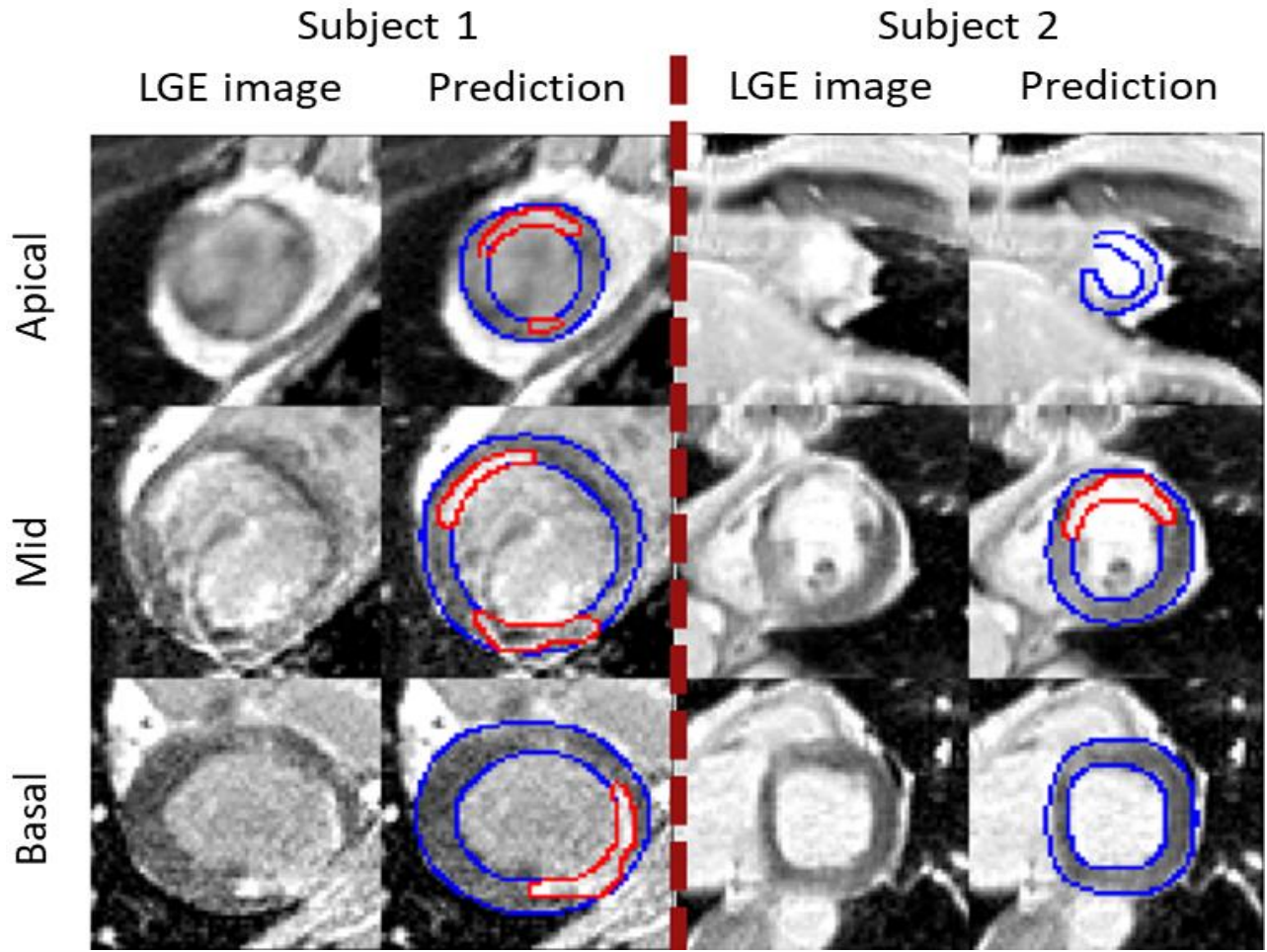

**Figure 5:** *The segmentations of two representative patients selected from the test set, showing the LGE image and the predicted scar (red) and myocardium (blue) segmentations. It can be seen that in subject 1 (left) and in the mid-slice of subject 2 (right) scar is accurately identified by the models. However, for subject 2, the myocardium segmentation in the apical slice is inaccurate. Note that the manual ground-truth for the test set is not publicly available.*

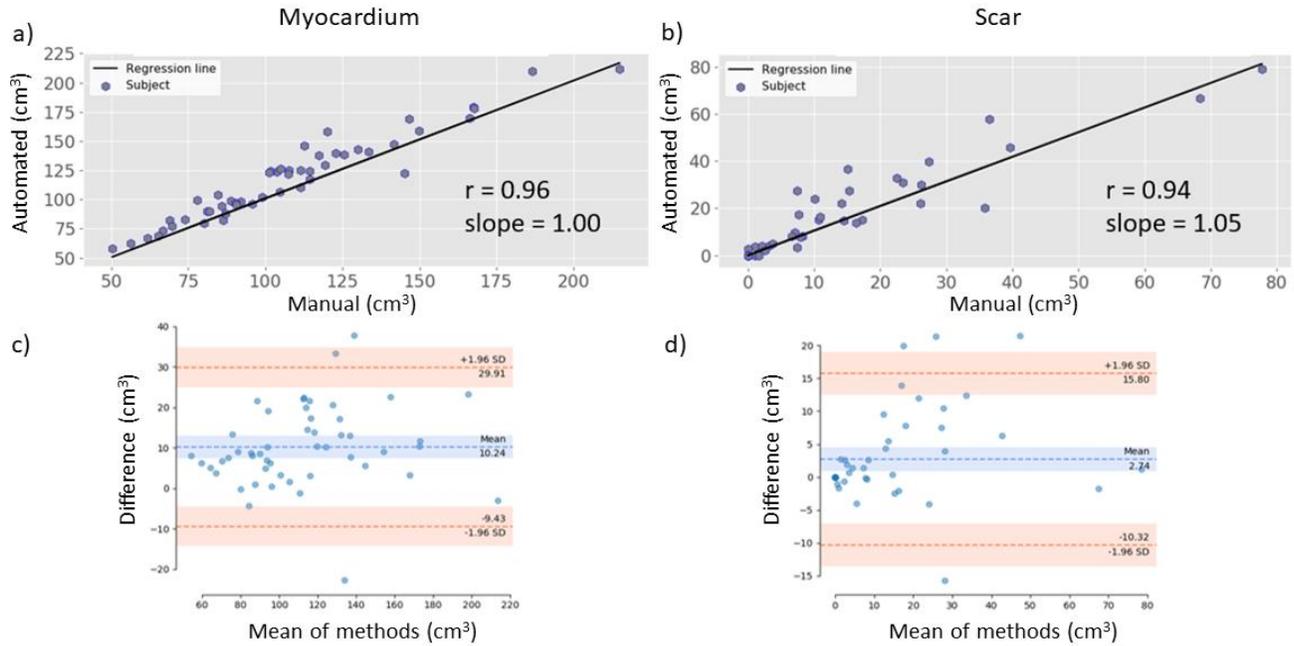

***Figure 6:*** *The top row shows a scatterplot of the manual versus automatic per-subject segmentation volumes for the myocardium and scar segmentation ((a) and (b)) with the Pearson correlation coefficient (r) and line of best fit, with the slope reported. (c) and (d) shows the Bland-Altman analysis for the myocardium volume (c) and scar volume (d).*

## 4    Discussion

In this work, two approaches are studied to learn from small datasets for the segmentation of scar from LGE cardiac MRI: the splitting of the task into smaller sub-problems and the use of synthetic data to increase the amount of available data. It is thought that the simpler sub-problems can be solved more effectively with the limited amount of available data and then applied in a cascaded pipeline to improve performance. In particular, a cascaded model was proposed that used three consecutive neural networks to identify the left ventricle, delineate the left ventricular myocardium and segment regions of myocardial infarction. The pipeline was trained based on manual segmentations of publicly available LGE cardiac MR images from the automatic Evaluation of Myocardial Infarction from Delayed-Enhancement Cardiac MRI (EMIDEC) challenge. Additionally, a segmentation-conditional GAN was proposed that uses SPADE layers coupled by a ResNet encoder to synthesize realistic LGE images on given augmented labels, for the purpose of data augmentation.

The proposed cascaded pipeline outperformed direct segmentation consistently in both the left ventricular myocardium and scar segmentation by a mean DSC increase of 0.21 per-slice on the internal test set. The three-step pipeline also improved over the combinations of two-step pipelines. Furthermore, the performance was improved by the synthetic image augmentation, with a mean DSC increase of 0.06 for the scar segmentation. These reported DSC results were comparable to the inter- and intra-observer variability found by Lalande et al. (20), inter-observer 0.84 and 0.76, intra-observer 0.83 and 0.69 for myocardium and scar segmentation, respectively. The impact of the synthetic data augmentation was also studied without the potentially confounding effects of the cascaded pipeline. That is, for the direct segmentation of scar and myocardium using a single 2D nnU-Net, models were trained with and without the synthetic data augmentation. As also found for the cascaded pipeline, for the direct segmentation the model trained with synthetic data augmentation significantly outperformed the model trained without the synthetic data.

A potential disadvantage of the cascaded approach is that errors can be propagated through the steps of the pipeline so that if, for example, an error is made in segmenting the myocardium, it will impact the subsequent scar segmentation. To test the effect of this error propagation, the model for scar segmentation was applied using the ground-truth myocardial segmentations and compared to using the predicted myocardium segmentations, on the internal test set. There is a small increase in mean DSC for the scar segmentation from 0.78 to 0.80 indicating that the negative impact of the cascaded pipeline is minimal and the cascaded approach still significantly outperforms the alternatives. This result confirms the findings of the original challenge, where cascaded pipelines were seen to perform well (28).

Our mean myocardium (0.86) and infarction (0.67) DSC scores compare favorably with the challenge results (28), with the combined DSC scores only being outperformed by a single participant. This winning solution of Zhang reported a mean DSC of 0.88 for the myocardium and 0.71 for the infarction regions (22). Zhang proposed a two-step system in which the coarse segmentation output of an initial 2D model was then input to a further 3D model to improve the 3D spatial consistency of the segmentations. For the purely 2D solutions, our proposed pipeline represents a new state-of-the-art, though it is shown that more sophisticated architectures still have the potential to improve on this. One of the typical disadvantages of using a 3D model in this application is that there are much less 3D images for training than 2D slices. However, as was shown in this work, it is possible to use synthetic images to augment the training dataset, and this is a possible future line of research to exploit the 3D nature of the data.

The current work uses rotations with dilation and opening of scar to augment the segmentation labels to input to the synthetic data generator. This work could be extended to use more complex patterns of scar and increase the robustness of the model. For example, patients with hypertrophic cardiomyopathy (HCM) often have complex patchy scar patterns and this could be simulated to allow training with a synthetic cohort of HCM patients without having to manually generate the training labels. Since the trained generator synthesizes the images based on a given LGE style image, different style images, from a difference acquisition sequence for example, could also be used to generate a more diverse training set.

The thorough evaluation of the impact of the synthetic data in this work using a challenging dataset with varying levels of contrast, noise, and artifacts indicates that the benefit is also likely to be more generally applicable to different applications and is also similar to that found in previous studies (29–31). In addition to the use of GAN-based synthetic data, an advantage can also be seen to a cascaded pipeline, where the overall task is split into more manageable sub-problems, and together these approaches can be exploited to lower the manual annotation burden of deep learning.

### 4.1 Limitations

The major limitation of this work is that it used a homogenous cohort of selected patients, from the EMIDEC challenge. These images were acquired at a single center, using scanners from a single vendor and uniform imaging protocols. Therefore, the trained model may not generalize to different clinical cohorts due to the "domain-shift" (the varying levels of signal, noise and contrast, differing scan planning, and diverse disease patterns). Although methods are being developed to account for this (32), the model would need to be tested on images from different patient cohorts, scanners, and centers prior to clinical deployment.

Our approach of using a 2D model treats each imaging slice independently and does not take advantage of the 3D relations between the slices. Indeed, we observe suboptimal performance in the apical (e.g Figure 5) and basal slices, and in future work, 3D or long axis information could be incorporated to constrain the segmentations to be spatially consistent. This may aid the segmentations, particularly in more challenging positions such as the apex and when the LV outflow tract is present in the basal slices.

This work segmented the scar and MVO regions as only one region of infarction and did not separate the MVO regions. The pipeline could be adapted, in future work, to consider the MVO regions separately. It also only focused on the identification of infarctions and the pipeline could also be extended beyond the segmentation to also classify patients' disease (33). This could potentially incorporate spatial information of the scar, as well as other relevant clinical biomarkers to improve the classification (28).

### 4.2 Conclusion

In a population of patients with suspected acute myocardial infarction, our results demonstrate that a cascaded deep learning-based pipeline trained with augmentation by synthetically generated data leads to myocardium and scar segmentations and quantitative volume values that are similar to the manual operator. The three-step cascaded pipeline was shown to significantly outperform direct segmentation with a mean DSC increase of 0.26 per slice. Additionally, the inclusion of GAN-based synthetic images as data augmentation further improved the performance and yielded a further mean DSC increase of 0.06 per-subject for scar segmentation.

**Conflict of Interest**

MB is an employee of Philips Healthcare. The authors declare that the research was conducted in the absence of any commercial or financial relationships that could be construed as a potential conflict of interest.

**Author Contributions**

**Funding**

This work was supported by the Wellcome/EPSRC Centre for Medical Engineering [WT 203148/Z/16/Z] and the openGTN project, supported by the European Union in the Marie Curie Innovative Training Networks (ITN) fellowship program under project No. 764465 (website: opengtn.eu).

**Acknowledgments**

The authors would like to thank the EMIDEC challenge organizers for the evaluation of the test set results.

## *Supplementary Material*

**Supplementary Figures**

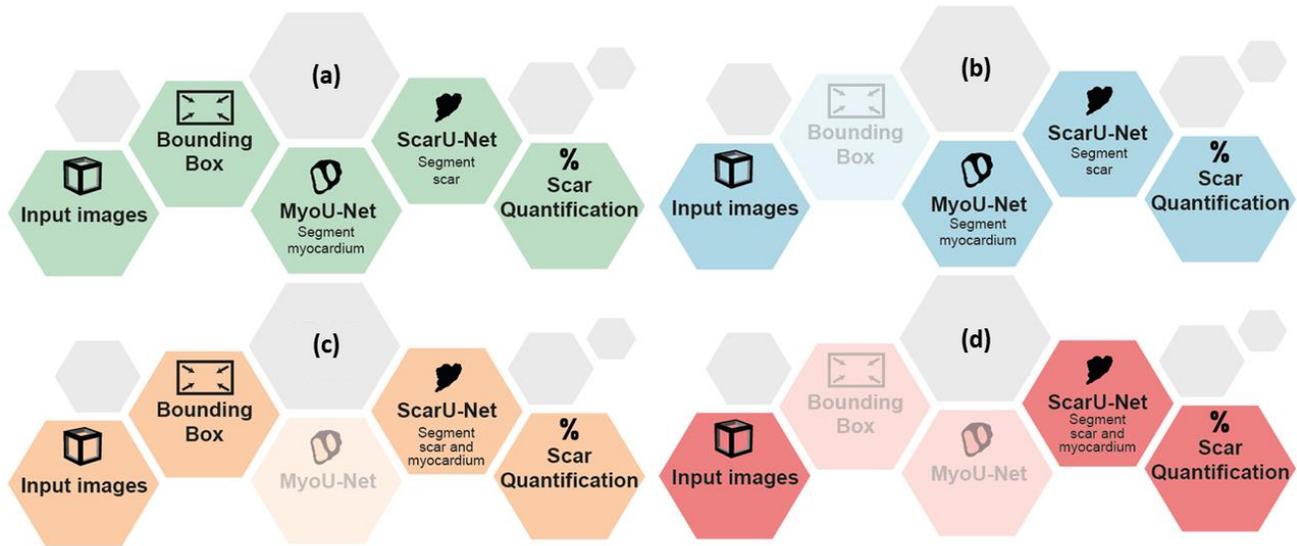

**Supplementary Figure 1.** A schematic representation of the ablation test, showing the different combinations of steps tested in the pipeline.

**Supplementary Tables**

**Supplementary Table 1.** The parameters used for the data augmentation in the bounding box training. $\mathcal{U}$(a, b) denotes that the parameter value was randomly sampled from a uniform distribution on the interval [a, b].

| Type of Augmentation | Value |
|---|---|
| Gaussian Noise | $\mu = 0.1, \sigma = 0.1$ |
| Gaussian Blur | $\sigma = 1.5$ |
| Shear | $\mathcal{U}$(-20, 20) |
| Rotation | $\mathcal{U}$(-90, 90) |
| Translation (independent in x and y direction) | $\pm\mathcal{U}$(0.14, 0.21) |
| Scale | $\mathcal{U}$(0.5, 1.5) |

**Supplementary Table 2:** The parameters used for the augmentation of ground-truth labels for image synthesis.

| Type of Augmentation | Value |
|---|---|
| Rotation | 0, 60, 120, 180, 240, 300 |
| Elastic deformation[1] | $\alpha = 50, \sigma = 5$ |
| Morphological opening | - |
| Morphological dilation | - |

---

[1]https://github.com/aleju/imgaug